\documentclass[useAMS,usenatbib]{mn2e}
\usepackage{amsmath} 
\usepackage{subfigure}
\usepackage{times}
\usepackage{graphicx}
\usepackage{times}
\usepackage{natbib}
\usepackage{times}
\def\Mpch{Mpc/{\it h}}
\def\Msun{M$_\odot$}
\def\Msunh{M$_\odot$/{\it h}}

\title[PopulationIII and LGRBs]
      {Population III stars and the Long Gamma Ray Burst rate}
\author[M.A. Campisi et al.]{
M.A. Campisi$^1$\thanks{E-mail: campisi@dfm.uninsubria.it},
U. Maio$^2$,
R. Salvaterra$^{1}$,
B. Ciardi$^{3}$\\
$^1$Dipartimento di Fisica e Matematica, Universit{\'a} dell'Insubria, via
Valleggio 7, 22100 Como, Italy\\
$^2$Max-Planck-Institut f\"ur extraterrestrische Physik,
Giessenbachstra{\ss}e 1, D-85748 Garching bei M\"unchen, Germany\\
$^3$Max-Planck-Institut f\"ur Astrophysik Physik,
Karl-Schwarzschild-Stra{\ss}e 1, D-85748 Garching bei M\"unchen, Germany
}
\begin{document}

\date{Accepted ???. Received ???; in original form ??? }
\pagerange{\pageref{firstpage}--\pageref{lastpage}} 
\pubyear{2011}

\maketitle

\label{firstpage}
\begin{abstract}
Because massive, low-metallicity population III (PopIII) stars may produce very powerful long gamma-ray bursts (LGRBs), high-redshift GRB observations could probe the properties of the first stars.
We analyze the correlation between early PopIII stars and LGRBs by using cosmological N-body/hydrodynamical simulations, which include detailed chemical evolution, cooling, star formation, feedback effects and the transition
between PopIII and more standard population I/II (PopII/I) stars. From the {\it Swift} observed rate of LGRBs, we estimate the fraction of black holes that will produce a GRB from PopII/I stars to be in the range 0.028$<f_{GRB}<$0.140, depending on the assumed upper metallicity of the progenitor.
Assuming that as of today no GRB event has been associated to a PopIII star, we estimate the upper limit for the fraction of LGRBs produced by PopIII stars to be in the range 0.006$<f_{GRB}<$0.022. 
When we apply a detection threshold compatible with the BAT instrument, we find that the expected fraction of PopIII GRBs (GRB3) is $\sim 10\%$ of the full LGRB population at $z>$6, becoming as high has 40\% at $z>$10.
Finally, we study the properties of the galaxies hosting our sample of GRB3. 
We find that the average metallicity of the galaxies hosting a GRB3 is typically higher than the critical metallicity used to  select the PopIII stars, due to the efficiency in polluting the gas above such low values.
We also find that the highest probability of finding a GRB3 is within galaxies with a stellar
mass $<10^7$~M$_\odot$, independently from the redshift.
\end{abstract}


\section{Introduction}\label{section1}

Gamma-ray bursts (GRBs) are the most energetic explosions in the Universe \citep{zhang04}, offering exciting possibilities to study astrophysics in extreme conditions and up to very high redshift ($z$), as shown by the detection of GRB~090423 at $z=8.2$ \citep{sal09b,tan09}.
According to recent studies and the so called ``collapsar model'', long GRBs (LGRBs), with durations longer than $2\,\rm{sec}$, are linked with the final evolutionary stages of massive stars.
In particular, a Wolf-Rayet star can produce a long GRB if its mass loss rate is ``small'', which is possible only when the stellar metallicity is very low.
Indeed, when metallicities are below $\sim 0.1-0.3\,Z_{\odot}$, the specific angular momentum of the progenitor allows the loss of the hydrogen envelope while preserving the helium core \citep{woo06b,Fryer_Woosley_Hartmann_1999}. 
Observations of high-$z$ LGRBs can provide unique information about the first
stages of galaxy formation in the early Universe and on the reionization process \citep{McQuinn09}. 
Moreover, it has been recently proposed \citep{mesz2010,toma10,suwa10} that also very massive, metal-free first stars, the so-called Population III (PopIII) stars, can produce a LGRB. 
The rotation rate has as well important implications for the evolution of the star and the possible production of GRBs. Indeed the stars should retain sufficient spin to power the collapsar burst \citep{bromm06,stacy2011,chiappini2011}.

Indeed, a recent study by \cite{suwa10} shows that the jet can potentially break out the stellar surface even if the PopIII star has a massive hydrogen envelope, thanks to the final long-lasting accretion of the envelope itself onto the central black hole (BH).
Since GRB luminosity and duration are found to be very sensitive to the core and envelope mass, they are probe of the first luminous objects at the end of the dark ages.
\cite{toma10} calculated the afterglow spectra of such PopIII GRBs based on the standard 
external shock model, showing that they could be detectable with the {\it Swift} BAT/XRT. 
Anyway, the rate of GRB from PopIII, even if measurable, is very small. Thus, observations of 
PopIII stars might be possible only with a large area and with the fields of view of future missions (as e.g. the Fermi LAT instruments).

Thus, GRBs associated to PopIII stars
should be observable out to redshift $\sim 30$ \citep[e.g.][and references therein]{ciardi00,toma10} and 
might represent the most promising way to directly detect the very first stars \citep{Lamb2000,bro02,salvaterra2010}.
\\
In this paper, we compare the redshift distribution of LGRBs from PopII/I stars with the one expected by PopIII stars up to high redshift, by using state-of-the-art N-body/hydrodynamical simulations \cite[see][]{MaioPhD,maio2010,maio2010arXiv,maio2011arX}, including detailed chemical evolution of the Universe, gas cooling, PopIII-to-PopII/I transition, and feedback effects.
To select the candidates for LGRBs,  we extract the information for the age and metallicity of newly formed stars, and adopt the collapsar model for both PopII/I and PopIII stars.
We build three samples of possible progenitors with different metallicity thresholds, and give estimates of the PopIII GRB rate measured by {\it Swift}.
We additionally compute the contribution of the LGRBs from PopIII progenitors with respect to the total LGRBs rate.
\\
The paper is organized as follows: in Sec. \ref{section2}, we present the N-body/hydrodynamical chemistry simulations used in this work; in Sec. \ref{section3}, we describe the method for selecting  different progenitors of LGRBs; we describe our results in Sec. \ref{section4}; then, we describe the main properties of the galaxies hosting our LGRBs from PopIII in Sec.\ref{host}; finally, we discuss results and give our conclusions in Sec. \ref{section5}.


\section{Numerical simulations}\label{section2}
We perform N-body/hydrodynamical simulations, including cosmological evolution of e$^-$, H, H$^+$, H$^-$, He, He$^{+}$, He$^{++}$, H$_2$, H$_2^+$, D, D$^+$, HD, HeH$^+$ \cite[][]{yoshida2003,maio2006,maio2007,maio2009}, PopIII and PopII/I star formation \cite[][]{tornatore2007} according to corresponding initial mass functions (IMF), gas cooling from resonant and fine-structure lines \cite[][]{maio2007} and feedback effects \cite[][]{springel2003}.
The transition from the PopIII to the PopII/I regime is determined by the underlying metallicity, $Z$, of the medium, compared to the critical value $Z_{crit}=10^{-4}Z_\odot$ \cite[see e.g.][and references therein]{Schneider2003,Bromm2003,schneider2006}.
If $Z<Z_{crit}$, then a Salpeter PopIII IMF is assumed with mass range between $100\,\rm M_\odot$ and $500\,\rm M_\odot$;
otherwise, a standard Salpeter IMF is adopted with mass range between $0.1\,\rm M_\odot$ and $100\,\rm M_\odot$, and SNII range between $8\,\rm M_\odot$ and $40\,\rm M_\odot$ \cite[see][]{maio2010arXiv,bromm09}.\\

The chemical model follows the detailed stellar evolution of each SPH particle. At every timestep, the abundances of the different species are consistently derived, according to the lifetimes of the stars and the yields in the given mass range\footnote{
Lifetimes are from \cite{PadovaniMatteucci1993} \cite[but see also, e.g.][]{MaederMeynet1988}. The yields used are from \cite{WoosleyWeaver1995} for SNII, \cite{vandenHoekGroenewegen1997} for AGB stars, \cite{Thielemann2003} for SNIa, \cite{heger2002} for PISN. For a more extensive and detailed discussion on different IMFs and metal yields and their effect on the enrichment history, see also \cite{maio2010} (in particular Fig. 6 and Fig. 7), and \cite{maio2011arX}(e.g. Fig. 5).
}.
Metal mixing is mimicked by smoothing the metallicities over the SPH kernel. Pollution is driven by wind feedback, which causes metal spreading over scales of several kpc at each epoch \cite[e.g.][]{maio2010arXiv}.
We warn the reader, though, that, due to poorly known quantities in stellar evolution modeling, different assumptions can determine variations in the PopIII-SFR estimates of more than one order of magnitude, as already shown in \cite{maio2010,maio2010arXiv}, and in \cite{maio2011arX}.
 More specifically, the largest effects are due to the PopIII IMF assumed in the simulations. Many studies suggest the PopIII IMF to be top-heavy, with stellar masses of the order of hundreds solar masses \cite[e.g.][]{Larson1998,abel2002,yoshida2004}.
However, there is also evidence for a more standard, low-mass PopIII IMF, with masses well below $\sim 100\,\rm M_\odot$ \cite[e.g.][]{Yoshida2006,Yoshida2007,Campbell2008,SudaFujimoto2010,greif2011}.
The main difference between these two scenarios is the lifetime of the first SN, which is $\sim 10^6\,\rm yr$ in the former case, and $\sim 10^8\,\rm yr$ in the latter, and this results in a transition from the PopIII to the PopII/I regime which takes place earlier in the top-heavy IMF scenario \cite[][]{maio2010,maio2011arX}.
On the other hand, uncertainties in the critical metallicity for the PopIII/PopII-I transition, $Z_{crit}$, are not expected to be crucial, since the rapid pollution process boosts metallicities up to $\sim10^{-3}Z_{\odot}$ in few $10^7$~yr after the first explosions \cite[][]{maio2010}.

In the simulation used here, the cosmological field is sampled at redshift $z=100$, with dark-matter and baryonic-matter species, according to the standard cosmological model, with total-matter density parameter at the present $\Omega_{0,m}=0.3$, cosmological constant density parameter $\Omega_{0,\Lambda}=0.7$, baryonic-matter density parameter $\Omega_{0,b}=0.04$, expansion rate in units of 100~km/s/Mpc $h = 0.7$, spectral normalization $\sigma_8=0.9$, and primordial spectral index $n=1$.
We consider three cubic volumes with comoving sides of 100~\Mpch, 10~\Mpch, and 5~\Mpch, with corresponding gas mass resolution of
$\sim 3\times 10^8\, {\rm M_{\odot}}/h$,
$\sim 3\times 10^5\, {\rm M_{\odot}}/h$,
$\sim 4\times 10^4\, {\rm M_{\odot}}/h$,
and dark-matter resolution of
$\sim 2\times 10^9\, {\rm M_{\odot}}/h$,
$\sim 2\times 10^6\, {\rm M_{\odot}}/h$,
$\sim 3\times 10^5\, {\rm M_{\odot}}/h$.\\
The identification of the simulated objects (with their gaseous, dark and stellar components) was carried out by applying a friends-of-friends technique, with comoving linking length of $20\%$ the mean inter-particle separation, and a minimum number of 32 particles. Substructures are identified by using a SubFind algorithm, which discriminates among bound and non-bound particles.
The parameters used in the simulations are listed in Table~\ref{tab:sims}.
\\
\begin{table}
\caption{Simulation parameters.}
\centering
\begin{tabular}{lccc}
\hline
Box side  & Gas particle & Dark matter particle & Maximum physical\\
{}[\Mpch] & mass [\Msunh]& mass[\Msunh]         & softening [kpc/{\it h}]\\
\hline
100 & $3\times 10^8$ & $2\times 10^9$ & 7.50\\   
10  & $3\times 10^5$ & $2\times 10^6$ & 0.50\\   
5   & $4\times 10^4$ & $3\times 10^5$ & 0.25\\   
\hline
\end{tabular}
\label{tab:sims}
\end{table}

\begin{figure}
  \centering
  \includegraphics[scale=0.45,angle=-90]{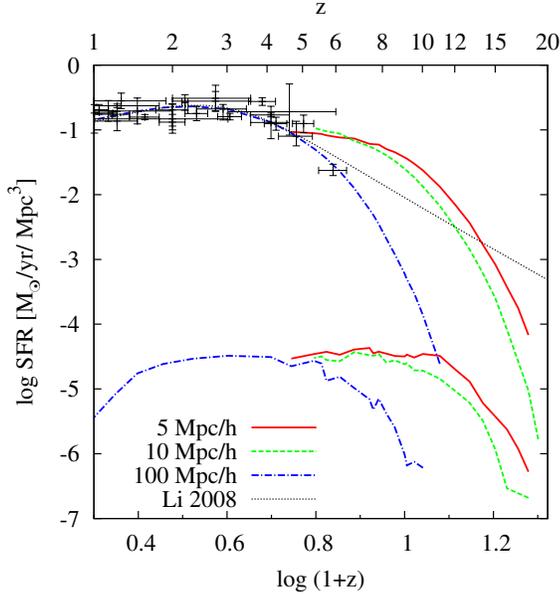}
\caption{Star formation rate density, $SFR\,[\rm M_{\odot}\,yr^{-1}\,Mpc^{-3}]$, as a function of redshift for the simulation 
with comoving side box of 5 Mpc/$h$ (red-solid line), 10 Mpc/$h$ (green-dashed line) and 100 Mpc/$h$ (blue-dotted-dashes line) . For
each of them, we show the total star formation rate density (upper lines) and the SFR of particles with $Z<Z_{crit}$ (PopIII SFRs),
i.e. that will produce a GRB3 (lower lines). Symbols with error bars are a compilation 
of observational data (Hopkins and Beacon 2006), the black-dotted line is the best fit for observational data (Li 2008b).
} 
\label{fig:sfr2}
\end{figure}


We show the analysis of the simulation outputs by reconstructing the evolution of the star formation rates (SFRs) in our simulations and comparing with observed data in Fig.\ref{fig:sfr2}.
We find that the total SFR of our larger simulation is in good agreement with the observational data,
while the smaller boxes were not evolved below $z \sim 5$ because at these redshifts most of the
power in structure formation is at larger scales. Nevertheless, we see that the total SFR converges
at $z \sim 5$. In addition, the general trend is very similar, with the contribution from 
PopIII rapidly decreasing with redshift and reaching $\sim 10^{-4}$ the total SFR.
Because of the higher resolution obtained, differently from the 100~Mpc/$h$ box, 
within the two smaller boxes, we marginally resolve the formation
of minihalos. Such primordial objects
have typical masses below $\sim 10^8\,\rm M_\odot$ and as small as $\sim 10^5\,\rm M_\odot$ and they
witness the formation of the first stars. As detailed e.g. in \citet{tornatore2007} and \citet{maio2010}, 
the first PopIII stars form 
in these small mass halos, of pristine composition, and their subsequent formation history depends
on the details of metal enrichment and chemical feedback.
For this reasons the smaller boxes are better suited to capture the star formation process at high redshift,
when, according to the standard hierarchical scenario of structure formation, the galaxy population is 
dominated by objects with masses much smaller than the present ones.

\section{Different Progenitors for GRBs}\label{section3}

We divide the possible progenitors in three subsamples: two associated with PopII stars ($Z > Z_{crit}$), with different upper cuts in metallicity, and one from PopIII stars ($Z\leq Z_{crit}$). 
As mentioned in Sec.~\ref{section1}, recent studies on the final evolutionary stages of 
massive stars suggest that a Wolf-Rayet star can produce a LGRB if its mass loss rate is small. 
This is possible only if the metallicity of the star is low, and therefore the specific angular momentum of the progenitor allows the loss of the hydrogen envelope while preserving the helium core \citep{woo06b,Fryer_Woosley_Hartmann_1999}.
The loss of the envelope reduces the material that the jet needs to cross in order to escape, while the helium core should be massive enough to collapse and power a GRB.\\
In order to select possible progenitors for LGRBs, we have extracted from the simulations the information about the age and the metallicity of all the star particles.
In particular, the metallicity is computed as the ratio between the total mass of metals contained in star
particles and the total stellar mass.
According to the collapsar model, we consider young star particles with age $t< t_c = 5\times 10^7 {\rm yr}$ \citep{Lapi_etal_2008,Nuza_etal_2007}. We follow an approach similar to
 \cite{campisi09}, and we will refer to our subsample in the following way: 
\begin{itemize}
\item
      GRB1, obtained by selecting star particles associated with PopII/I stars ($Z > Z_{crit}$), without upper metallicity cut;
\item
      GRB2, including star particles associated with PopII/I stars ($Z > Z_{crit}$), and with metallicity $Z\leq0.5Z_{\odot}$;
\item
      GRB3, defined by selecting star particles with metallicity $Z\leq Z_{crit}$. 
\end{itemize}

In particular, for the second subsample we adopt a metallicity cut of 
$\sim0.5\,Z_{\odot}$ because stronger metallicity cuts seem to be excluded
by the analyses of the properties of observed GRB host galaxies at $z<1$ \citep{mannucci2010,campisi2011}.
\\
In the following, we will briefly describe how we compute the rate of LGRBs (Sec.~\ref{sect:GRBrate}), considering both PopII/I (Sec.~\ref{sect:popIIGRB}) and PopIII (Sec.~\ref{sect:popIIIGRB}) progenitors.

\subsection{LGRBs rate}\label{sect:GRBrate}

The formation rate of LGRBs in the simulations can be computed by:
\begin{equation}
\rho_{GRB,i}(z)=f_{GRB,i}\,\,\zeta_{BH,i}\,\,\rho_{*,i}(z)
\label{eq:1}
\end{equation}
where $i$ indicates the 3 subsamples previously described, $f_{GRB,i}$ is the fraction of BHs that 
will produce a LGRB, $\zeta_{BH,i}$ is the fraction of BHs formed per unit stellar mass,
and $\rho_{*,i}(z)$ is the star formation rate for the considered subsample, $i$.
\\
The observed rate (in units of [yr$^{-1}$ sr$^{-1}$]) of LGRBs of the $i$-th 
subsample at redshift larger than $z$ is then given by  
\begin{eqnarray}
\label{number}
R_{GRB,i} (>z)&=&\gamma_b\int_z dz' \frac{dV(z')}{dz'} 
\frac{1}{4\pi} \frac{\rho_{GRB,i}(z')}{1+z'} \nonumber \\
& \times & \int_{L_{th}(z')} dL^\prime \psi_i(L^\prime),
\label{eq:2}
\end{eqnarray}
where $\gamma_b$ is the beaming fraction, $dV(z)/dz$ is the comoving 
volume element, and $\psi_i$ is the normalized LGRB luminosity function (LF). 
The factor $(1 + z)^{-1}$ accounts for the cosmological time dilation. 
The last integral gives the fraction of LGRBs with isotropic equivalent peak 
luminosities $L>L_{th}$, i.e. those LGRBs that can be actually detected by the satellite.
The threshold luminosity $L_{th}$ is obtained by imposing a 1-s peak photon
flux limit of $P=0.4$ ph s$^{-1}$ cm$^{-2}$ in the 15-150 keV band of 
{\it Swift}/BAT.
\\
There is a general consensus that GRBs are jetted sources \citep{wax98, rho97}
implying fundamental corrections to their energy budget and GRB rates. We
take into account this fact by considering that the observed LGRB rates should
be corrected by the factor $\gamma_b$ that is related to the jet opening angle
$\theta$ by $\gamma_b = (1-\cos\,\theta) \sim \theta^2/2 $ \citep{sar98}. Given the 
average value of $\theta\sim 6^{\circ}$ \citep{ghi07}, 
$<\gamma_b> \sim 5.5\times10^{-3}$, which is the value we adopt.
\\
Finally, in order to compute the observed rate we have to specify the LF of LGRB, 
$\psi_i(L)$, for the three LGRB subsamples considered here. Although the
number of LGRBs with good redshift determination has been largely increased
by {\it Swift}, the sample is still too poor (and bias dominated) to allow
a direct measurement of the LF. However, an estimate of the LGRB LF and of its
evolution through cosmic times can be obtained by fitting the BATSE 
differential number counts and imposing the constraint on the bursts observed 
by {\it Swift} (e.g. \cite{sal07,sal09,campisi2010}, but see \cite{Butler2010,Wanderman2010}
for different approaches).
\\
In particular, for GRB1-2 subsamples we adopt the LF obtained by 
\cite{campisi2010}, who described the LF as 
\begin{equation}
\psi(L) \propto\left ( \dfrac{L}{L_*} \right ) ^{\xi} \exp \left (-\dfrac{L_*}{L} \right ),
\end{equation}
\noindent
where $\xi$ is the bright-end power index and $L_*$ is a characteristic 
cut-off luminosity. In order to take into account possible evolution of the 
LGRB LF, the cut-off luminosity can be written as 
$L_*=L_0(1+z)^{\sigma}$ where $L_0$ is the cut-off luminosity at $z=0$.
Since the SFR evolution of our GRB1 and GRB2 samples are similar to the 
Host1 sample in \cite{campisi2010}, we adopt here for both subsamples the 
following value for the LF free parameters: $L_0=0.3\times 10^{50}$ erg 
s$^{-1}$, $\sigma=2$ and $\xi=-1.7$, that provide a good description of 
the available data. For GRB3, we assume that the LGRB from PopIII stars
are expected to be much brighter than PopII/I LGRB. 
Since typical luminosities should exceed $10^{53.6}$ erg s$^{-1}$ \citep{toma10},
we adopt $L_*=10^{54}$ erg s$^{-1}$ constant in redshift and $\xi\sim -1.7$. 
However, we consider a range of possible values with $L_*=10^{53}-10^{55}$  
erg s$^{-1}$ and $-1.5<\xi <-2.0$. We will discuss
the dependences of our results on these parameters in Sec.~\ref{section4}.

\subsection{LGRBs from PopII/I stars}\label{sect:popIIGRB}
\begin{table}
\centering
\caption{Fraction of BHs that will produce $GRB_i$, $f_{GRBi}$.
For $i$ =1,2 two different values for the minimum stellar mass to die as a BH are considered.
For $i$=3 two different values for the number of GRBs observed by {\it Swift}, $N_{GRB_{Swift}}$,
are used to constrain $f_{GRB3}$ . See text for details.
}

\begin{tabular}{|lcc|}
\hline
\hline
$f_{GRB_i}$    &from BHs with &from BHs with \\
& $M>20M_{\odot}$ & $M>40M_{\odot}$\\
\hline
$f_{GRB1}$& $\sim0.028$&$\sim0.089$\\
$f_{GRB2}$& $\sim0.044$&$\sim0.140$\\
\hline
\hline
&\multicolumn{2}{c}{$N_{GRB_{Swift}}$}\\
&500 (up) &140 (up2)\\
\hline
$f_{GRB3}$& $\sim0.006$ &$\sim0.022$ \\
\hline
\hline
\end{tabular}
\label{tab:2}
\end{table}

The LGRB formation rate (see definition in Eq.~\ref{eq:1}) can be obtained from the 
star formation rate in the $i$-th subsample once $\zeta_{BH,i}$ and $f_{GRB,i}$
have been specified. The fraction of BHs formed per unit stellar mass can 
be computed by 
$\zeta_{BH,i} =\int_{m_{min}}^{100} \phi(m_*) dm_*/\int_{0.1}^{100} m_* \phi(m_*) dm_*$
where $\phi(m_*)$ is the Salpeter IMF and $m_{min}$ is the minimum stellar mass
of stars dying as BHs. Given the uncertainties on $m_{min}$ we will consider
here two cases with $m_{min}=20\,\rm M_\odot$ and $m_{min}=40\,\rm M_{\odot}$, the latter consistent with the value assumed in the numerical simulations.\\
Since not every BHs produces a LGRB, as described in eq.~\ref{eq:1}, we need 
to know the efficiency with which this happens, $f_{GRB,i}$. This can be estimated
 by considering the observed rate of LGRBs detected by {\it Swift}. 
Indeed, as of today the {\it Swift} rate of LGRBs with redshift $z>1$ is about\footnote{
Here, we assume that the fraction of LGRBs observed at redshift $<1$ is $\sim20\%$ of the whole sample of bursts, as estimated in \cite{campisi2010}.}
$57$ yr$^{-1}$ sr$^{-1}$.
By imposing this value, we find $f_{GRB,1} = $0.028 (0.089) and $f_{GRB,2} = $0.044 (0.140),  assuming $m_{min}=20\,\rm M_\odot$  ($m_{min}=40\,\rm M_\odot$).
These values correspond to having $\sim7$ ($\sim 11$) LGRBs every
1000 SNe for the GRB1 (GRB2) subsample.
We point out here that to compute the $f_{GRB,i}$, we use the simulation with box sizes 100 Mpc/$h$, since this is the one reaching lower redshifts. We then assume the same value of $f_{GRB,i}$ also for the other simulations.
The adopted values are summarized in Table~\ref{tab:2}.

\subsection{LGRBs from PopIII }\label{sect:popIIIGRB}

In a similar way, we have to specify the value of $\zeta_{BH,3}$ and $f_{GRB,3}$ for GRB3.
As already mentioned, the first stars have been assumed to be very massive, with masses $>100$ M$_\odot$.
Stars with masses between $140-260$~M$_{\odot}$ are expected to die as pair-instability supernovae \citep[][]{heger2002,Zeldovich1971,Zeldovich1999}, leaving no compact remnant.\\
Thus, only progenitors with masses in the range $100-140 \rm M_{\odot}$ and $260-500 \rm M_{\odot}$ will lead to the formation of a BH possibly triggering a LGRB event.
For the assumed PopIII IMF we obtain $\zeta_{BH,3}\sim$0.0032.\\
\\
The fraction of PopIII stars that will produce LGRBs is unknown.
As of today, there is no strong evidence favoring the detection
with {\it Swift} (or any other satellite) of a LGRB associated to a PopIII star. Indeed,
even the most distant LGRB detected so far shares common prompt emission and
afterglow properties with the low- and intermediate-$z$ LGRB population 
\citep{sal09b}, suggesting that it likely originates from a normal
PopII progenitor. Thus, we can set a firm upper limit on the fraction of
PopIII stars triggering the LGRB event by imposing that no LGRB associated
with a PopIII star has been detected by {\it Swift}\footnote{Since the simulations
stops at $z=1$, when calculating $R_{\rm GRB3}(>0)$ we assume that the SFR from 
PopIII stars is negligible at $z<1$.}, that is:
\begin{equation}
 \frac{R_{\rm GRB3}(>0)}{R_{\rm GRB}(>0)}< \frac{1}{N_{Swift}},
\label{stimaf}
\end{equation}
\noindent
where $N_{Swift}$ is the number of LGRBs observed by {\it Swift}.
 Thus an upper limit can be set by considering no PopIII 
LGRB in the whole sample of {\it Swift} bursts. At the time
of writing this consists in 500, corresponding to $f_{\rm GRB3_{up}}=0.006$.
A more gentle constraint can be obtain by considering that some PopIII LGRB
can be hidden among those {\it Swift} bursts without redshift measurement. In
this case,  $f_{\rm GRB3_{up2}}=0.022$ is obtained by imposing no PopIII GRB
present among the 140 LGRBs with measured $z$ (Table~\ref{tab:2}). 
These values correspond to have 1 GRB3 every $<2 (7) \times 10^{5}\,M_{\odot}(Z<Z_{crit})$ produced.
We note that a part of the high-$z$ GRBs population might be missed by
current follow up observations, as about 20\% of all bursts appear as ``dark'',
possibly hiding some high-$z$ GRBs. However, there is increasing observational
evidence that most GRBs are ``dark'' because of dust extinction \citep{cenko09,greiner2011}.

\begin{figure*}
  \centering
  \includegraphics[scale=0.45,angle=-90]{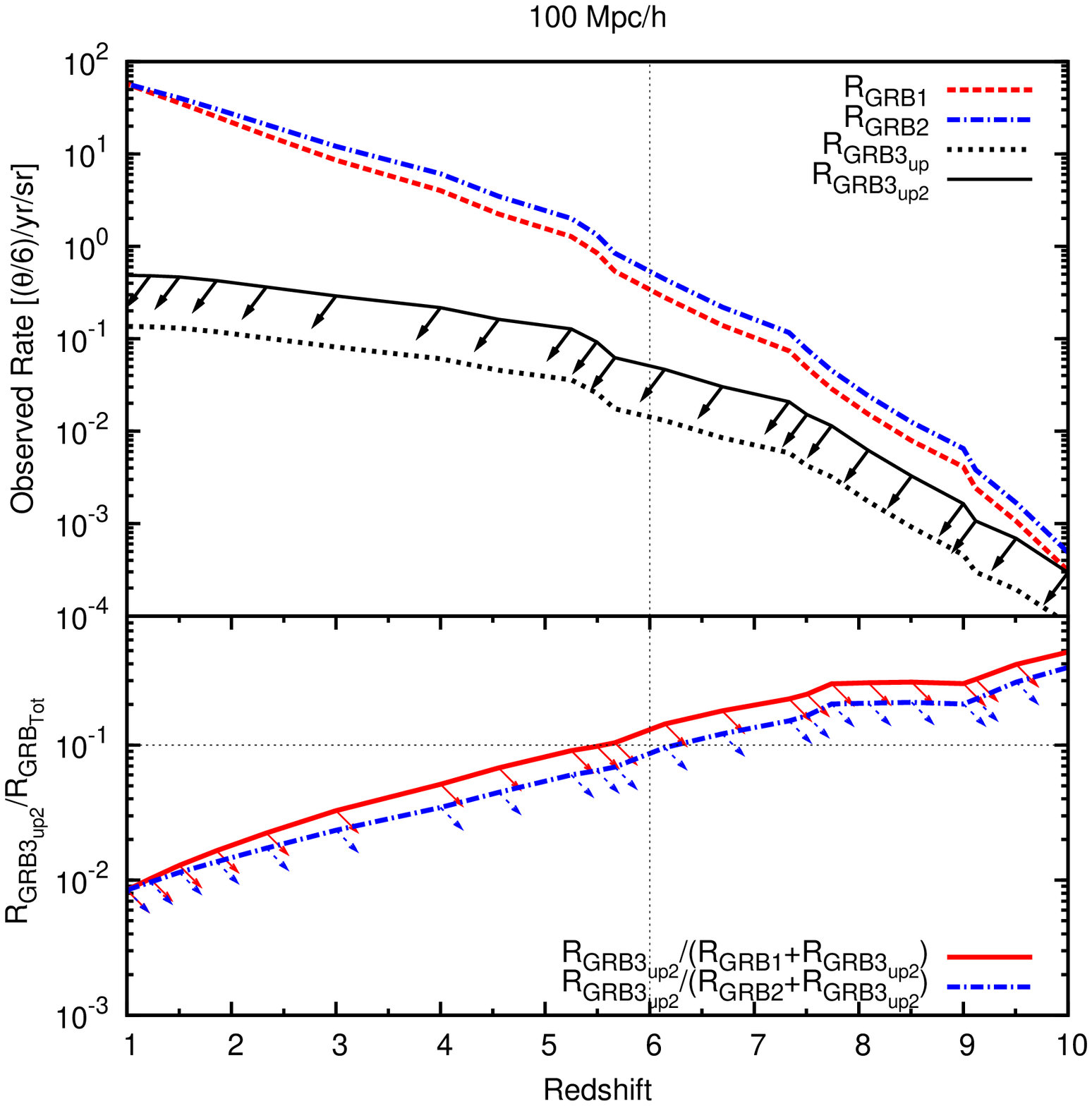}
  \includegraphics[scale=0.45,angle=-90]{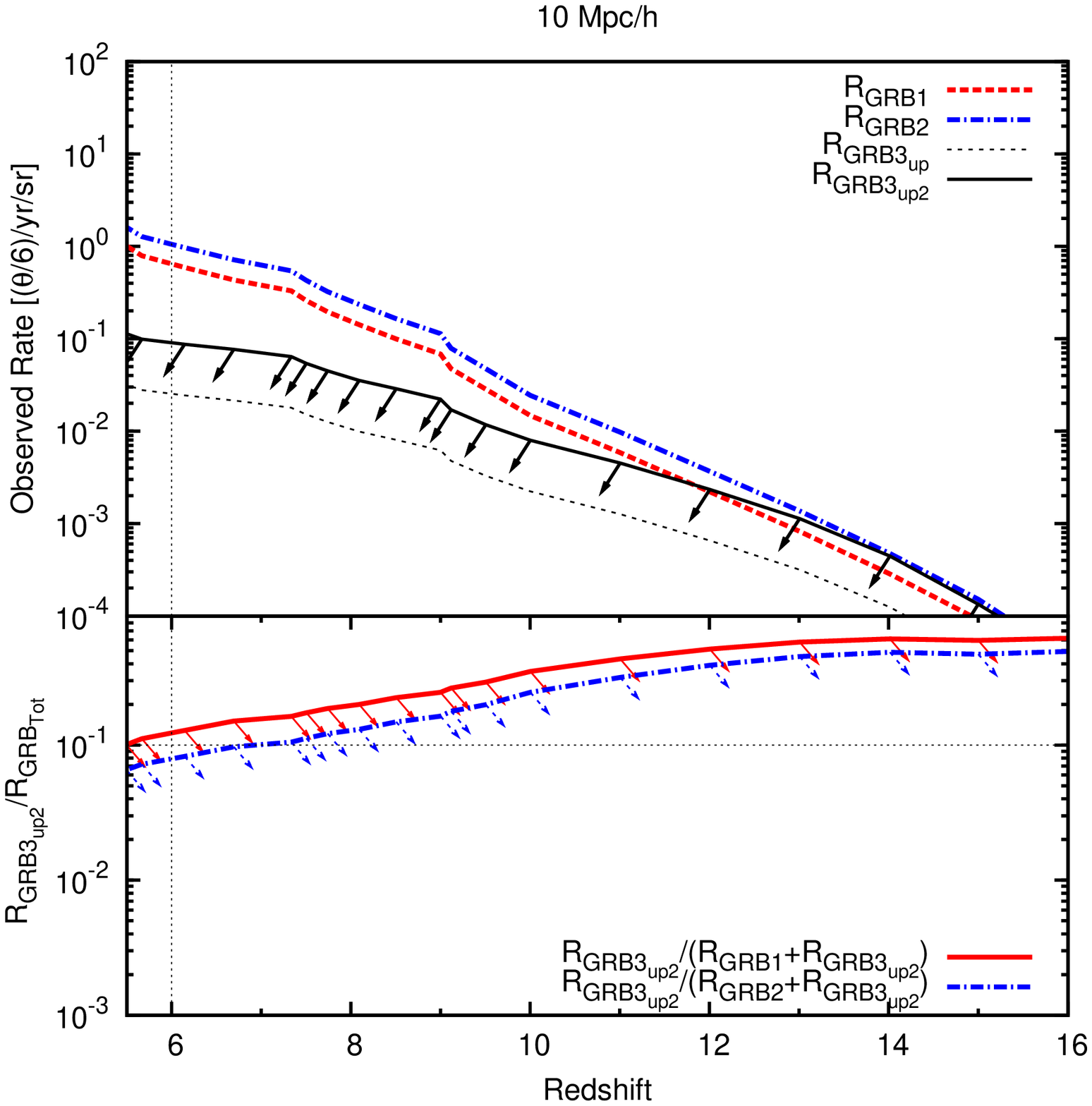}
  \includegraphics[scale=0.45,angle=-90]{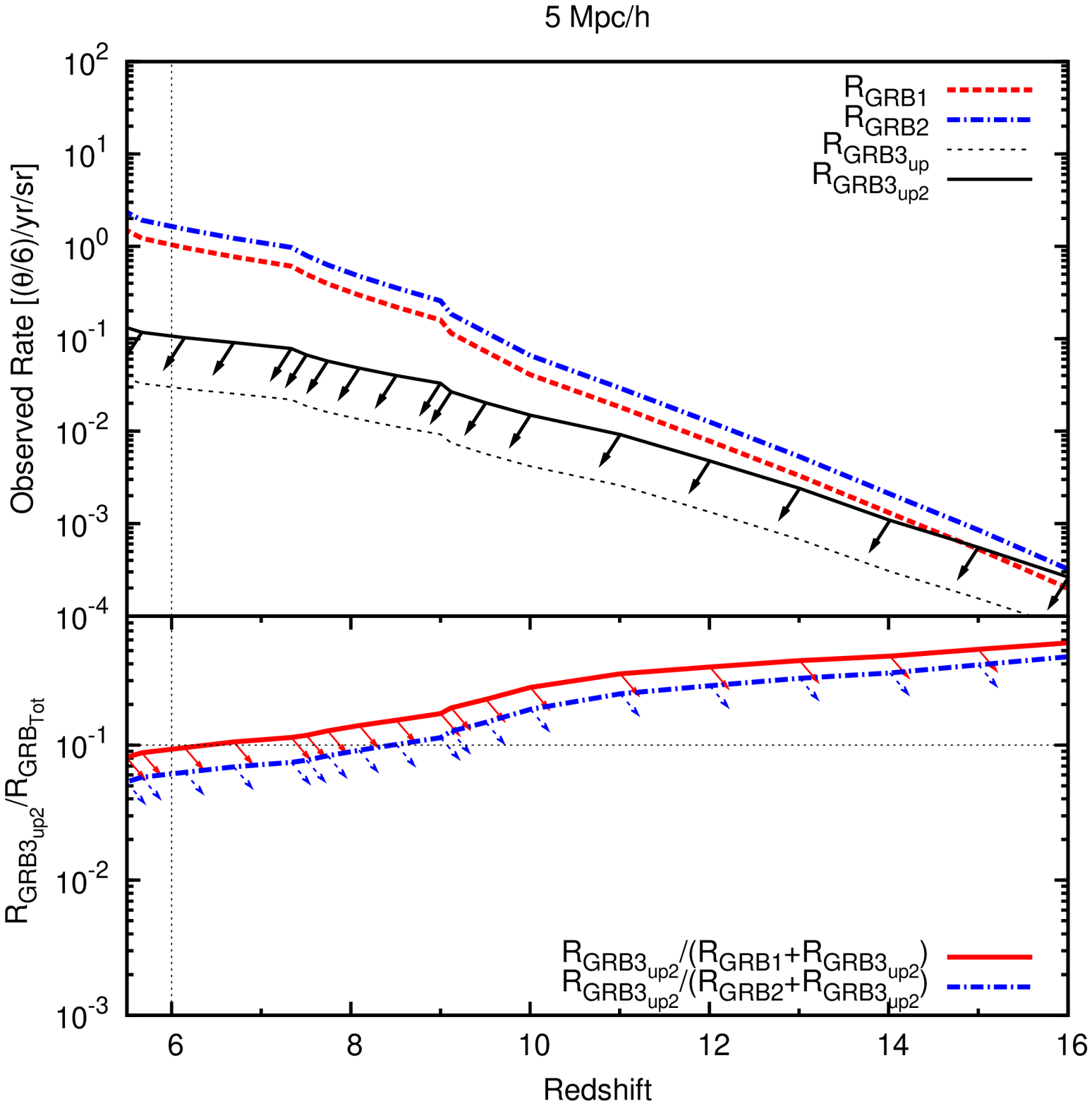}
  \caption{{\it Top Panels}: Observed rate for our samples of LGRBs in the three simulations with different side box. 
Dashed-red lines represent the rate of GRB1, dashed-dotted-blue lines the rate of GRB2, the solid-dark lines are the upper limits for the rate of GRB3 and dotted lines are the more restrict upper limit for the same sample (using $f_{\rm GRB3_{up}}$ and $f_{\rm GRB3_{up2}}$ respectively). 
{\it Bottom Panels}: Evolution with redshift of the ratio between the expected rate of GRB3 and the
 total rate obtained summing the rate of the samples GRB3+GRB1 (upper solid-red line) and GRB3+GRB2 (lower dashed-dotted-blue line), for the three simulations with different size box. 
}
\label{fig:rate1}
\end{figure*}
\section{Results}
\label{section4}

The redshift distributions of LGRBs of different subsamples obtained by 
integrating eq.~\ref{number}  are shown in Fig.~\ref{fig:rate1}.
Here we plot the evolution with redshift of observed cumulative rate for our 
subsamples of LGRBs in simulations with 
different box size. 
Dashed-red lines represent the rate of GRB1 ($R_{\rm GRB1}$), while dashed-dotted-blue lines
 the rate of GRB2 ($R_{\rm GRB2}$). Note that the different
choice of the minimum mass of the BHs does not alter the redshift evolution
of LGRBs. 
The solid-dark line is the upper limit for $R_{\rm GRB3}$ case, obtained using $f_{\rm GRB3_{up2}}$,  and dotted-line is the rate for the same sample computed using $f_{\rm GRB3_{up}}$. We use the arrows to point out that the computed rate of GRB3 are upper limits. In the bottom panel of each plot we 
report the ratio between the rate of GRB3 and the total rate, i.e. the
sum of the rate of PopII/I GRBs and PopIII GRBs. 
The solid line refers to the case in which LGRB from PopII/I
stars comes from GRB1 subsample whereas dot-dashed line to GRB2 subsample. 
Here we assume $f_{\rm GRB3_{up2}}$.
\\
If we consider the total rate of LGRB for the 100~Mpc/$h$ simulation, i.e. including the contribution
from both PopII/I and PopIII stars, the expected observed rate at redshift $z>$6 is
 about $R_{\rm tot}(z>6) \sim 1\, (\theta/6)^2 \rm /yr/sr$.
This is consistent with recent observational constraints \citep{perley09,greiner2011},
 as well as with previous theoretical estimates \citep{sal07,sal09,Butler2010}. 
At these redshifts, the corresponding rate of GRB3 is below 0.017 (0.06) sr$^{-1}$ yr$^{-1}$
for $f_{\rm GRB3_{up}}$ ($f_{\rm GRB3_{up2}}$), 
and is consistent with \cite{salvaterra2010} when the same value of $f_{\rm GRB3}$ is adopted.
At $z>6$ the expected fraction of PopIII  LGRBs is $\leq 10\%$ and increases with redshift.
However, at  $z\sim8$ it can not be larger than 20\%.
 This is in agreement with a PopII progenitor for GRB~090423 \citep{sal09b,chandra2010}.
\\
To extend our results to very high $z$ we turn now to our smaller box simulations.
Indeed, as already discussed in Sec.~2, at these redshifts the cosmic SFR is dominated by low-mass objects that can be properly resolved only by our small-box simulations.
We note that at $z\sim 5-6$ the results of the smaller simulations converge and are 
in good agreement with those obtained from the 100~\Mpch{ } side box run. 
As expected, the contribution of PopIII GRBs is more and more important as redshift increases, becoming dominant for $z >16$.
Thus, a LGRB detected at extremely high redshift would likely be originated from a PopIII star.
\\
These results are not very sensitive on the assumed LF for PopIII LGRB. 
In fact, we find that the rate of PopIII LGRBs at $z>6$ is within a factor of two if we vary the characteristic burst luminosity in the range $10^{53}-10^{55}$ erg s$^{-1}$ and the slope in the range $-1.5<\xi< -2.0$.

\begin{figure*}
  \centering
  \includegraphics[scale=0.8,angle=-90]{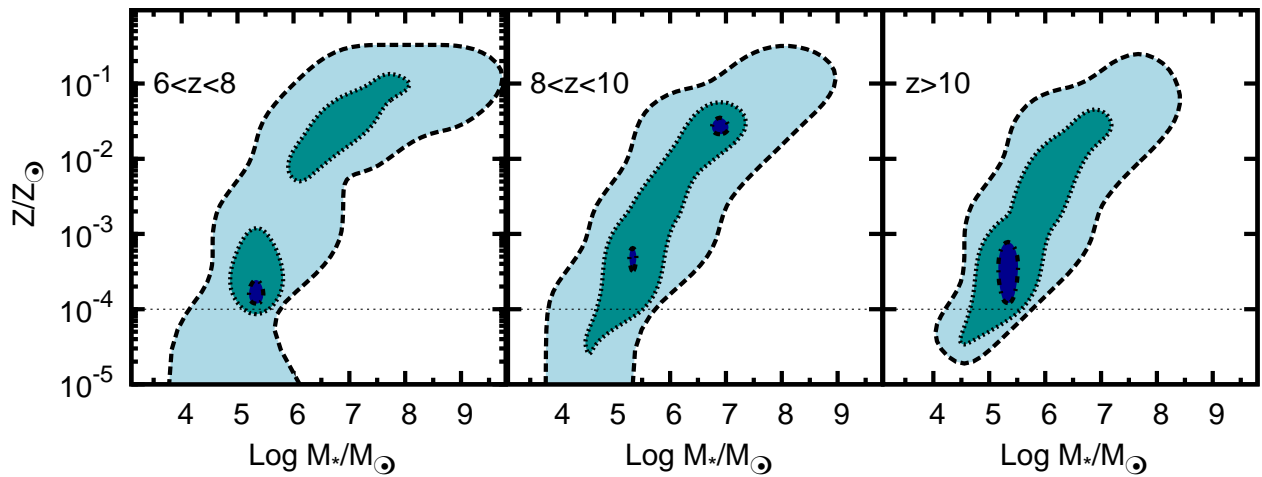}
  \caption{Probability of having a GRB3 in a host galaxy with given stellar mass and metallicity, in three
different redshift bins. The contours refer to a probability of 100\%, 75\% and 25\%, where the percentages refer to the contours from the outermost to the innermost, respectively. The horizontal line
indicates the critical metallicity $Z_{crit}$.
}
\label{fig:host}
\end{figure*}

\section{host galaxies of GRB3 sample}\label{host}

In Sec.~\ref{section2} we describe how the host galaxies are obtained from the simulations. 
The number of expected LGRBs from PopIII in the $k$-th galaxy can be 
computed as $N_{k,3}=f_{GRB,3}\,\,\zeta_{BH,3}\,\,M_{*,3,k}(z)$, where $M_{*,3,k}$
is the stellar mass in the $k$-th galaxy that satisfies our selection
criteria for the $3$-rd subsample (see Sec.~\ref{section3}). 
Here we are interested in characterizing the properties of PopIII GRB hosts
at high $z$.
To this end, we consider the results obtained in the 10~Mpc/$h$ side box run
(this choice is motivated by the fact that we need to resolve properly
the PopIII hosts and to have sufficient statistics to compute average quantities).
We then calculate the probability of having a GRB3 in a host galaxy of a given stellar mass,
$M_*$, and metallicity, $Z$, in different redshift bins.\\
In Fig.~\ref{fig:host} we show this probability, weighted for the total number of LGRBs, present in the full galaxy sample.
Most of the simulated GRB3 hosts are expected to be found in a very
well defined region of the $M_*-Z$ plane. These galaxies have always values of $Z$ lower than the 
solar metallicity at all redshift, but typically higher than the critical value used to select PopIII stars 
($Z_{crit}=10^{-4} \,Z_{\odot}$).
This is because metal enrichment from massive primordial stars proceeds rapidly, and, as soon as 
they form within a galaxy, they quickly pollute the surrounding environment to values above 
$Z_{crit}$ \cite[see e.g.][]{wise08,greif2010,maio2010,maio2011,maio2011arX}.
Despite this, pockets of metal free gas are found within metal enriched galaxies 
\cite[see e.g.][]{tornatore2007}, where PopIII star formation can occurr. The details, though, depend
on the underlying model of metal enrichment and on the level of turbulent mixing which cannot be
properly resolved by cosmological simulations.
\\
GRB3 also seem to reside typically in objects at the lower end of the stellar mass distribution, $M_* \sim 10^5$~M$_\odot$, because these have a higher probability of hosting non polluted gas. On the other hand, they should rarely be found in objects with $M_* > 10^7$~M$_\odot$.
It should be noted that this result may depend on the resolution of the simulation. We expect in fact,
the typical stellar mass of the host galaxy to decrease when minihalos are properly resolved. \\

\section{Discussions and conclusions}\label{section5}

LGRBs are now detected up to extreme high redshift and are considered 
promising tool to study the state of the Universe during and beyond the
reionization epoch. Moreover, it has been recently argued that even very
massive, metal-free first stars could in principle trigger a LGRB event \citep{woo06b,Fryer_Woosley_Hartmann_1999}.
 Given the expected relation between LGRB properties and the
BH mass, PopIII LGRBs are predicted to be very bright and detectable with
current and future instrument up to $z>20$ \citep{bromm06,naoz2007,toma10}. If this is the
case, the detection of PopIII LGRBs might represent
the most promising way to directly probe the nature of the very first stars \citep{salvaterra2010}.
\\
In this work, we have used three N-body/hydrodynamical cosmological simulations, with various box sizes ($L$=100 Mpc/$h$ - 10 Mpc/$h$ - 5Mpc/$h$), to constrain 
the number and redshift distribution of PopIII LGRBs and compare them to those
of PopII/I LGRBs. 
To estimate the rate of LGRBs of various populations,  we assume the collapsar model, and construct three samples of LGRBs: GRB1 are bursts produced by PopII/I stars, GRB2 produced by PopII/I stars with metallicity $Z<0.5Z_{\odot}$ and GRB3 from PopIII stars ($Z<10^{-4}\,Z_{\odot}$).
In order to compute the rate of PopII/I LGRBs at high $z$, we follow an observationally motivated approach, 
by adopting a LGRB luminosity function able to reproduce both the BATSE differential number counts
and the redshift distribution of LGRBs detected by {\it Swift} \citep{campisi2010}.
We find that $\sim 1$~yr$^{-1}$~sr$^{-1}$ PopII/I LGRBs are predicted to lie at $z>6$,
consistently with available observational constrains \citep{perley09,greiner2011}
 as well as with most recent theoretical works \citep[e.g.][]{sal09,Butler2010,Wanderman2010}. 
In order to constrain the
contribution of PopIII LGRBs, we compute an upper limit on the fraction of
PopIII stars that can power a LGRB, assuming that none of the bursts detected
by {\it Swift} up-to-now comes from a PopIII progenitor. With this limit, we
predict that less than 0.06 sr$^{-1}$ yr$^{-1}$ PopIII LGRBs should be detected
at $z>6$. PopIII LGRBs are found to be non-dominant at $z<10$ with respect
to normal PopII/I LGRBs. However, as expected, their contribution increases with
redshift and the probability to find a PopIII LGRBs at  $z>8$
is $\sim$20\%, reaching also $\sim40-50\%$ at higher redshift, when the contribution
of PopIII stars to the total star formation rate is maximum.
This implies that if we observed a LGRB at redshift $z>10$, it would have roughly the same probability to be produced by a PopII/I or PopIII star.
\\
Taking advantage of the detailed description of the simulated galaxies,
we further investigate the properties of the possible hosts of 
PopIII-LGRBs at very-high redshift. We find that the average metallicity of the
galaxies hosting a GRB3 is typically higher than the critical metallicity used to select the PopIII stars, due to the efficiency in polluting the gas above such low values.
We also find that the highest probability of finding a GRB3 is within galaxies with a stellar
mass $<10^7$~M$_\odot$, independently from the redshift.
\\
They key assumption we have made throughout this paper is that the primordial PopIII IMF is top-heavy with mass range between 100\Msun{} and 500\Msun.
However, there are many uncertainties about that and different studies \cite[e.g.][]{Yoshida2006,Yoshida2007,Campbell2008,SudaFujimoto2010} show that smaller-mass first stars could be born even at very early times. This would imply longer stellar lifetimes and two main predictable consequences \cite[see][]{maio2010,maio2011arX}:
(i) a general delay in the cosmic enrichment history of roughly $\sim 10^8\,\rm yr$;
(ii) a larger contribution to the PopIII star formation rate at high redshift.\\
It is not the aim of the paper to investigate this alternative scenarios, but if this were true, as a consequence we would expect a corresponding increment in the GRB3 rate at high $z$.
\\
Thanks to the simultaneous observations from the {\it Swift} and {\it Fermi} satellites \citep{bissaldi11}, future
detections of high-$z$ GRBs might help discerning the contribution from PopIII stars, and
if at least 10 GRBs will be detected at $z>6$ one should be a GRB3. At the present, though, the number of
such high-$z$ GRBs is too small and the available data not good enough to discriminate their origin.
Nevertheless, the probability of detecting a high-$z$ GRB3 is non-negligible, and it becomes comparable
to the probability of detecting a GRB from PopII/I stars at $z>10$.
Thus, investigation of high-$z$ GRBs with future missions (e.g. SWOM, EXIST, ORIGIN, Janus)
 could offer a realistic first opportunity to detect PopIII stars.


\section*{Acknowledgements}
The authors thank Luca Graziani and Klaus Dolag for important discussions, help and suggestions on  numerical aspects of the paper.
They also thank the referee, Volker Bromm, for his useful comments which helped improving the presentation of the paper.
The simulations and the data analyses were performed on the AMD Opteron machines of the Garching computing center (Rechenzentrum Garching, RZG) of the Max Planck Society.
For the bibliographic research we made use of the tools offered by the NASA ADS archive.

\bibliographystyle{mn2e}
\bibliography{paper}
\label{lastpage}

\end{document}